# Development and Benchmarking of a Blended Human-AI Qualitative Research Assistant

JOSEPH MATVEYENKO, RAND Corporation
JAMES LIU, RAND Corporation
JOHN DAVID PARSONS, RAND Corporation
RYAN A. BROWN, RAND Corporation
ALINA PALIMARU, RAND Corporation
PRATEEK PURI, RAND Corporation

Qualitative research emphasizes constructing meaning through iterative engagement with textual data. Traditionally this human-driven process requires navigating coder fatigue and interpretative drift, thus posing challenges when scaling analysis to larger, more complex datasets. Computational approaches to augment qualitative research have been met with skepticism, partly due to their inability to replicate the nuance, context-awareness, and sophistication of human analysis. Large language models, however, present new opportunities to automate aspects of qualitative analysis while upholding rigor and research quality in important ways. To assess their benefits and limitations - and build trust among qualitative researchers - these approaches must be rigorously benchmarked against human-generated datasets. In this work, we benchmark Muse, an interactive, AI-powered qualitative research system that allows researchers to identify themes and annotate datasets, finding an inter-rater reliability between Muse and humans of Cohen's $\kappa$ = 0.71 for well-specified codes. We also conduct robust error analysis to identify failure mode, guide future improvements, and demonstrate the capacity to correct for human bias.

## 1 INTRODUCTION

Qualitative research has long been characterized by its emphasis on interpretive depth, contextual understanding, and the careful construction of meaning from textual data [17, 20]. The methodological foundations of qualitative analysis - from grounded theory to thematic analysis [7, 8, 11] - prioritize the researcher's interpretive lens, cultural sensitivity, and iterative engagement with data as essential components of rigorous inquiry [35]. Yet these same methodological strengths present significant practical challenges [9, 40]. The process of qualitative coding requires substantial time investment, with researchers often spending weeks or months in iterative cycles of reading, coding, discussing, and refining their analytical frameworks. Multiple rounds of inter-rater reliability (IRR) checks, consensus-building discussions, and theoretical refinement are standard practice to manage coder fatigue and interpretative drift [26], reflecting the field's commitment to analytical rigor [31] but also creating resource constraints that limit the scale and scope of many studies.

As qualitative research datasets grow in size and complexity with the exploitation of digital, online, and crowdsourced data [36], these resource constraints have become increasingly acute. Researchers face difficult tradeoffs between analytical depth, time, and costs - often opting to collect smaller samples or focus on narrow aspects of their data to maintain methodological rigor [10]. At the same time, many within the qualitative research community have long faced difficulties securing funding for their work - challenges that are likely to intensify as support for the social sciences, humanities, and other qualitative-research-driven fields continues to decline - sparking renewed interest in methodological innovations that preserve analytical rigor while reducing resource demands.

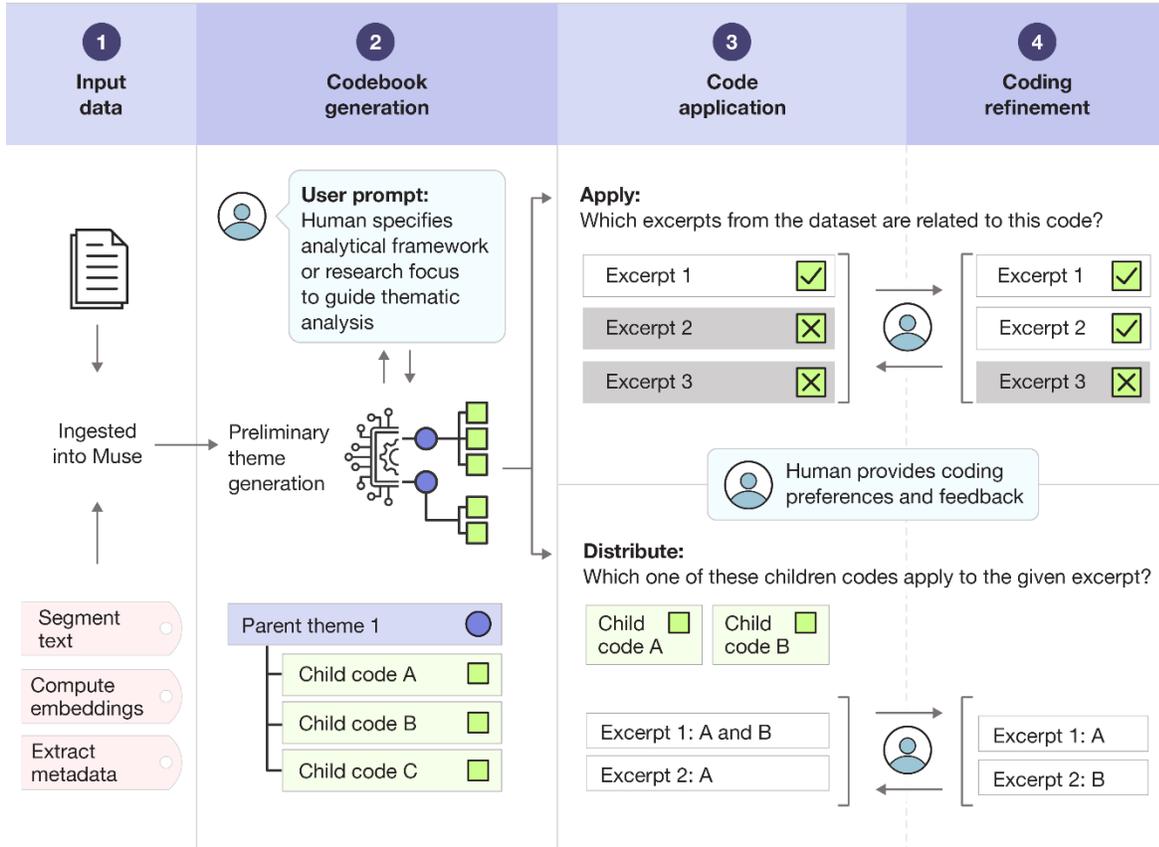

Figure 1: An overview of the Muse system for AI-assisted qualitative research. Documents are ingested into the platform, at which point they are segmented, embedded, and catalogued. Users may prompt the system to identify themes and generate a codebook. Similarly, users can apply a code to the entire dataset or distribute a set of excerpts amongst a series of codes. In all cases, the researcher directly interacts with the system to provide feedback, offer guidance, and otherwise align research outputs with their analytical framework.

In response to these pressures, some researchers have begun exploring computational approaches to augment qualitative analysis [14, 19, 46]. These explorations, however, have at times been met with considerable methodological skepticism within the qualitative research community [12, 50]. Concerns about losing interpretive nuance, introducing algorithmic bias, or undermining the fundamentally human process of meaning-making have led many researchers to view automated methods as potentially contaminating or undermining the analytical process [15, 47]. This tension reflects a deeper epistemological question: can computational tools assist qualitative analysis while preserving the interpretive depth and contextual sensitivity that define rigorous qualitative inquiry?

Recent advances in large language models (LLMs) have shown promise for automating aspects of qualitative analysis [3, 4, 29, 38]. For example, prior work, like LLooM [30], has demonstrated that AI can extract high-level concepts from unstructured text and generate interpretable topic models that surpass clustering-based approaches like BERTopic [21] in certain research contexts. However, a critical gap remains: these novel approaches need to be more rigorously benchmarked against real-world, human-generated datasets to assess their performance benefits, limitations, and trustworthiness in qualitative research. Compiling such datasets has been challenging, both because (1) many qualitative datasets contain sensitive personal information and thus are not suitable for public release and (2) there is a remarkable diversity in how qualitative research is performed, recorded, and framed, meaning standardizing research inputs and outputs across research projects can take considerable amounts of effort.

While qualitative research shares some core similarities with more rigorously-studied methods for analyzing text, such as text classification (comparable with code application) and topic modeling (comparable with codebook generation), there are several nuances within qualitative research that require more careful consideration and customized benchmarking. For example, rigorous code application requires a clear grasp of the research question, broader context in which a statement was made, and researcher intent. Relying entirely on standard natural language processing benchmarks to improve human-AI qualitative research workflows may not adequately address these factors, leading to potentially ineffective outcomes.

In this work, we perform comprehensive benchmarking on the Muse platform (Figure 1), an AI-powered qualitative research assistant that interacts with researchers to perform core qualitative research tasks while maintaining the rigor and transparency expected in this line of research. We explore how Muse can address three key challenges: (1) consistent application of established coding schemes across diverse datasets, (2) automated codebook generation that captures the nuances of qualitative frameworks, and (3) rigorous evaluation of AI-assisted coding quality using established IRR metrics. Through our analysis, we present the following contributions to the human-computer-interaction (HCI) community:

*Curated multi-domain evaluation dataset:* We compile and clean eleven publicly available qualitative research datasets spanning interviews, social media posts, survey responses, and domain-specific corpora. This collection addresses critical data sharing limitations in qualitative research and provides a foundation for standardized benchmarking of AI-assisted qualitative research, a much-needed area of inquiry.

*Systematic hyperparameter optimization for qualitative AI systems:* Through comprehensive evaluation across multiple datasets, we identify optimal configurations for LLM-based coding systems, tuning over parameters such as prompt design, model selection, and dataset batching. We explore the performance and cost/latency tradeoffs associated with different system settings, ultimately identifying parameters that judiciously balance both.

*Steerable codebook generation with empirical grounding*: We present algorithms that enable researchers to guide the development of automated codebooks from raw datasets through natural language instructions while maintaining empirical anchoring to source data. This approach allows for the rapid exploration of multiple analytical frameworks while preventing hallucinated themes unsupported by evidence.

*Error taxonomy and failure mode analysis*: We systematically categorize the root causes of AI coding errors in a representative qualitative dataset, identifying two primary failure modes: under-specified codes and insufficient document context. Informed by these findings, we provide best practices researchers can adopt to improve AI performance along with concrete design recommendations for future human-AI qualitative research systems. Additionally, we demonstrate the capability of LLM-based coding systems to serve as an audit trail for human coding teams or individuals to correct for human bias.

*Human-AI collaboration insights for qualitative research*: We demonstrate how AI assistance can achieve human-level coding reliability (Cohen's $\kappa$ = 0.71) for well-specified codes across a wide range of datasets and domains, comparable to performance demonstrated in domain-specific evaluations [6]. In addition, we highlight how Muse enables researchers to systematically both simultaneously explore multiple theoretical frameworks and validate coding consistency at scale, actions that are resource-constrained in many qualitative research projects.

## 2 SYSTEM DESIGN

To ensure our system addresses real-world research needs rather than technical benchmarks alone, we conducted several expert interviews with intended users and conducted a survey of over 100 experienced qualitative researchers. These touchpoints informed critical design decisions from feature prioritization to interface design to hyperparameter selection. Participants came from a wide range of methodological backgrounds, topic areas of expertise, and levels of AI familiarity. By grounding our technical development in the perspectives and opinions of this diverse set of researchers, we aimed to design a system that will generalize across research contexts and address researcher pain points in ways pure algorithmic optimization cannot.

When asked about high-priority tool features in our survey, researchers expressed a desire for flexible workflows that enable both inductive and deductive analysis, greater consistency in code application, and the ability to identify code definition issues that may compromise inter-rater reliability.

Table 1. Muse Functions

| Function Name | Description | Human Interaction Components |
| --- | --- | --- |
| generate_codebook | Conducts thematic analysis on text segments in a dataset to iteratively create and validate a hierarchical codebook | User may specify their research lens in natural language and their preferred level of topic granularity. These specifications guide an LLM toward identifying a codebook tailored to the researcher's need |
| apply_code | Uses a code definition, inclusion/exclusion criteria, and examples to assess its applicability to chunks of every document in the dataset | User can specify code definitions and inclusion/exclusion criteria before initial LLM coding. Additionally, the user can tune confidence score thresholds (1-10) after coding is complete to adjust the specificity by which a code is applied. Any false positives and false negatives identified through user interaction are fed back into the system to improve performance. |
| distribute_code | Assesses the applicability of chunks of every document to a list of all codes in a codebook or all children of a parent code | User specifies both whether multiple codes can be applied to the same excerpt and if at least one code must be applied to each excerpt. False positives and false negatives are identified through user interaction with an initial sample and fed back into the system to improve performance. |

Given these perspectives, we chose to focus on the development and evaluation of three main functions within the Muse platform that align with these needs. The *generate_codebook* feature generates a thematic coding structure from the raw transcripts associated with a qualitative research project. The *apply_code* function applies an existing code across a user's entire dataset by considering each excerpt in the dataset and assessing its relevance to a given code. Lastly, *distribute_code* determines which code(s) within a set of codes apply to a given document or excerpt, performing a similar function to *apply_code* with greater time efficiencies. Table 1 compares the functionality and human interactive aspects of these three main functions.

While the Muse platform contains many other features — such as chat-with-your-data functionality, automated code definition updating, and others — in this work, we focus primarily on the features in Table 1 given their amenability to rigorous evaluation and their general relevance to others building human-computer interfaces for qualitative research.

## 3 EVALUATION

In this section, we evaluate the performance of the *generate_codebook, apply_code,* and *distribute_code* functionalities within Muse. We find that codebook generation matches near-state-of-the-art clustering methods in accuracy while offering increased steerability and that code application achieves coding reliability comparable to humans (Cohen's $\kappa = 0.71$ for well-specified codes).

Table 2. Datasets Used to Evaluate Muse

| Dataset Name | Description | Data Type | Documents | Words | Codes[a] | Quality[b] |
|---|---|---|---|---|---|---|
| Reuters-21578 [32] | Reuters newswire documents | News | 18,918 | 2,446,793 | 135 | 6/10 |
| Medical Abstracts [45] | Medical abstracts of patient conditions | Medical | 11,550 | 2,074,849 | 5 | 7/10 |
| arXiv Categories [44] | arXiv titles and abstracts derived from metadata | Academic papers | 163,168 | 24,135,250 | 142 | 9/10 |
| Topic Annotations on Reddit Posts [41] | Eating disorders and dieting forums | Social media | 1,080 | 135,246 | 15 | 6/10 |
| Interviews on Data Curation [34] | Dissertation about qualitative reuse | Semi-structured interviews | 29 | 200,000 | 238 | 5/10 |
| Data for Search Systems Study [23] | Study about systematic searching and systems | Semi-structured interviews | 12 | 62,663 | 83 | 4/10 |
| Genocide Transcript Corpus [43] | Transcripts from three genocide tribunals | Legal | 52,845 | 1,782,057 | 1 | 7/10 |
| Wikipedia Discussion Corpora [18] | Wikipedia talk pages about improvements | Online discussion | 1,763 | 126,690 | 21 | 4/10 |
| Catalonia Independence Corpus [52] | Spanish and Catalan-language Tweets | Social media | 12,074 | 361,630 | 3 | 4/10 |
| Automated Hate Speech Detection [16] | Tweets containing hate speech lexicon | Social media | 24,783 | 349,862 | 2 | 6/10 |
| ATLAS.ti Sustainability [1][c] | Set of interviews on sustainability | Semi-structured interviews | 30 | 12,908 | 65 | 8/10 |

[a] Codes are the number of unique human-coded labels and themes in each dataset.
[b] Quality is a 1-10 dataset quality score that we assign based on an unweighted average of our assessment of codebook definition completeness, annotation consistency and document completeness.
[c] ATLAS.ti dataset used only to evaluate codebook generation.

### 3.1 Datasets for Evaluating System Performance

We used eleven publicly available human-coded datasets to assess the performance of Muse in generating codebooks and annotating qualitative data. These include Reuters-21578, a canonical dataset for text classification, two other text classification datasets containing medical and scholarly article abstracts, and eight datasets generated specifically within qualitative research projects. These datasets cover a diverse range of both data formats (social media, interviews, etc.) as well as domain areas (disordered eating, political tension, sustainability, etc.).

Because open-source qualitative datasets are typically scattered across diverse repositories and institutional websites with limited discoverability, we believe our systematic identification and compilation of the eleven datasets described in Table 2 is a useful contribution to the field. These datasets are in no means perfect; qualitative coding can produce artifacts and inconsistencies due to coder fatigue, drift, and code under-specification. Each researcher's coding scheme is unique and could require both high-level project context and days of training and discussion to replicate. These limitations may constrain the reliability of our assessments. However, even collecting these datasets is a starting point for the field, and the methodology discussed later in this report will only become more useful as higher quality datasets are collected. To provide a rough reference point for the relative quality of each dataset, we assigned a 1-10 dataset quality score based on an unweighted average of our assessment of each dataset's codebook definition completeness, annotation consistency, and document completeness.

Appendix A.1 describes our search strategy for identifying datasets, the rubric used to create the dataset quality score, and the preprocessing approach.

### 3.2 Generate Codebook Methodology

Our approach to automated codebook generation employs a multi-stage process that systematically analyzes qualitative data through progressive abstraction and refinement. The algorithm first segments the document set into units of $N_{LC}$ words (with a single segment capable of spanning multiple documents). A line-coding LLM ($LLM_{LC}$) is instructed to identify between $n_{min}$ and $n_{max}$ themes within each segment, with both values set by the user but defaulted with values identified within our survey. Specifically, $LLM_{LC}$ is prompted to both identify emergent codes and extract their operational definitions and supporting quotations from the data. The user can also provide natural-language instructions to $LLM_{LC}$ that will guide it towards discovering concepts aligned with a particular research scope or framing, or the user can opt not to provide any instructions, leading to more inductive codes. The resulting preliminary codebooks from each segment are then consolidated through a recursive codebook condensation process using a second LLM ($LLM_{CC}$). This consolidation occurs iteratively, merging sets of preliminary codebooks until a unified coding framework emerges that represents themes present across all data segments. This distributed coding approach ensures that themes are identified consistently across the entire corpus while preventing any single portion of the data from dominating the analytical framework.

The consolidated codebook undergoes validation and structural refinement to ensure both empirical grounding and analytical coherence. The system generates parent themes by prompting another LLM ($LLM_{PC}$) to (1) examine the relationships between the identified child codes in the previous set (2) identify a set of parent concepts that encapsulate these codes and (3) assign each child code to one of the identified parent codes. Again, the user can provide instructions to $LLM_{PC}$ to produce a more deductively defined parent theme hierarchy or opt out of providing such instructions.

Each code is then validated through an iterative process. First, we create a composite embedding for each code by averaging the embeddings of its definition and all associated quotations, then use Maximum Marginal Relevance (MMR) scoring to identify the top 20 most relevant chunks from the dataset. An LLM ($LLM_{CV}$) evaluates whether each code is sufficiently supported by these representative excerpts, and codes lacking adequate empirical grounding are removed from the codebook.

Finally, the algorithm performs hierarchical refinement by prompting another LLM ($LLM_{HR}$) to examine each parent code and its children to (1) merge semantically redundant codes that capture overlapping concepts and (2) split overly broad codes into more specific subcodes under appropriate general categories.

This computational approach mirrors the process of traditional line-by-line coding while leveraging natural language processing to achieve consistency and scalability across large qualitative datasets. The incorporation of researcher-specified theoretical frameworks and analytical priorities throughout the process helps ground the automated analysis within the study's epistemological stance and research objectives.

## 3.3 Generate Codebook Evaluation

We recognize that codebook generation is inherently subjective - a single dataset can be meaningfully mapped to many different analytical frameworks depending on research goals and theoretical perspectives. While the true value of codebook generation algorithms is best assessed through user studies, we developed a quantitative evaluation process to help optimize the numerous hyperparameters in our system during development.

In the first parts of this section, we detail this quantitative benchmarking. Afterwards, to demonstrate the practical benefits of our approach beyond these technical optimizations, we demonstrate how our algorithm leverages user-defined research preferences to generate tailored codebooks - a capability that distinguishes our LLM-based approach from many clustering-based topic modeling methods.

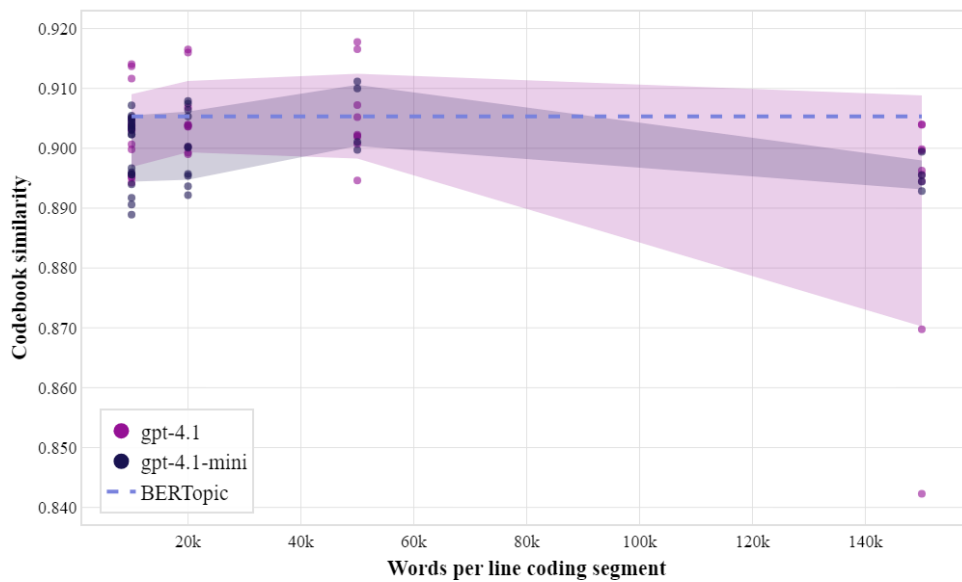

Figure 2: Codebook similarity as a function of the number of words within each line-coding segment for both gpt-4.1 and gpt-4.1-mini models. Similarity stays relatively constant as segment length is varied, demonstrating that larger length dataset segmentation can be used with lower-latency (gpt-4.1-mini models) can be used to produce codebooks that are similar in quality to those offered by near-state-of-the-art clustering methods such as BERTopic (dashed line). The banded regions represent 1σ intervals.

### 3.3.1 Dataset and Sampling

We benchmarked codebook generation against the arXiv Categories dataset [44], which provides an ideal evaluation context given its hierarchical codebook structure, large scale, and vetted quality. While not strictly qualitative data, arXiv abstracts contain the conceptual complexity and domain-specific language characteristic of many qualitative datasets.

We randomly selected three ground-truth parent codes from the arXiv corpus and drew 1,000 abstracts from documents labeled with at least one of those parents. For this 1,000-abstract sample, we included every child code associated with those documents, yielding a sample-specific codebook containing the selected parents and all sampled descendants. We repeated this procedure five times (with different parent-code triplets), producing evaluation dataset codebooks ranging from 22 to 52 codes - a range aligned with the typical codebook size reported in our internal survey.

*3.3.2 Similarity Metric for Codebook Comparison*

Evaluating algorithmically-generated codebooks presents unique challenges, as human coders can legitimately derive radically different frameworks from identical data. We developed a customized composite similarity metric for this purpose. With more details provided in Appendix A.2, our customized metric compares human- and LLM-generated codebooks by evaluating both semantic similarity (cosine embeddings of labels) and structural similarity (hierarchical depth and organization). These scores are computed for each pairwise combination of codes from each codebook, aligned using the Hungarian algorithm [27], weighted appropriately, and finally averaged to yield a blended similarity score that balances content and structure.

*3.3.3 Performance Optimization Results*

We systematically varied the ($N_{LC}$, $LLM_{LC}$, $LLM_{CC}$) hyperparameters and calculated codebook similarity scores across all five sample datasets under each setting. The highest-performing setting outperformed the lowest-performing setting by only ~3%, indicating the method is largely insensitive to these hyperparameters and shows only minimal accuracy–latency trade-offs. As shown in Figure 2, when selected as $LLM_{LC}$, both gpt-4.1 (higher-latency) and gpt-4.1-mini (lower-latency) exhibit minimal degradation as $N_{LC}$ grows from 20k to 140k words, an important finding given larger $N_{LC}$ values map to fewer LLM calls by extension lower algorithmic latencies. Because LLM-based theme generation is generally more compute-intensive than clustering, this hyperparameter robustness has meaningful implications for operationalization.

We established a comparison baseline (invariant across $N_{LC}$) by using BERTopic, an industry standard neural topic modeling framework. Our BERTopic implementation incorporated text-ada-002 embeddings, UMAP dimensionality reduction, and HDBSCAN clustering, with hyperparameters optimized via Bayesian optimization [5] with a customized objective function designed to balance coherence scores as well as user-specified codebook size constraints ($n_{min}$, $n_{max}$).

Our LLM-based method achieved comparable performance to the optimized BERTopic baseline (~0.90-0.91 similarity), demonstrating that our approach matches near-state-of-the-art clustering methods in accuracy while offering additional capabilities.

*3.3.4 Steerability of Algorithm*

The critical advantage of our approach lies in its steerable flexibility - researchers can direct codebook generation using natural language instructions to explore different analytical frameworks within the same dataset. This capability is difficult to achieve with clustering-based methods, where cluster formation is determined by optimization metrics which are often not straightforward to map onto research intent.

Figure 3 demonstrates this steerability by applying our codebook generation algorithm to the Atlas.ti Sustainability Dataset under three different user-specified research framings:
- *Causal factors*: "Identify the causal factors that shape individual views of sustainability and the environment"
- *Market dynamics*: "Focus on themes describing how market dynamics and consumerism impact individual views on sustainability"
- *Psychological barriers*: "Our research goal is to identify the psychological barriers that prevent sustainable action from being taken by individuals and communities"

The generated codebooks differ in research framing, semantic content, and conceptual organization, yet all remain grounded in the source data. This demonstrates how researchers can iteratively refine their analytical lens while maintaining empirical validity. In addition, our algorithm incorporates built-in safeguards against LLM hallucination through ground-truth anchoring. All generated codes must be supported by actual excerpts from the dataset, preventing the creation of theoretically plausible but empirically unsupported themes. This constraint ensures that even highly steerable and flexible codebook generation remains rigorously tied to the evidence base.

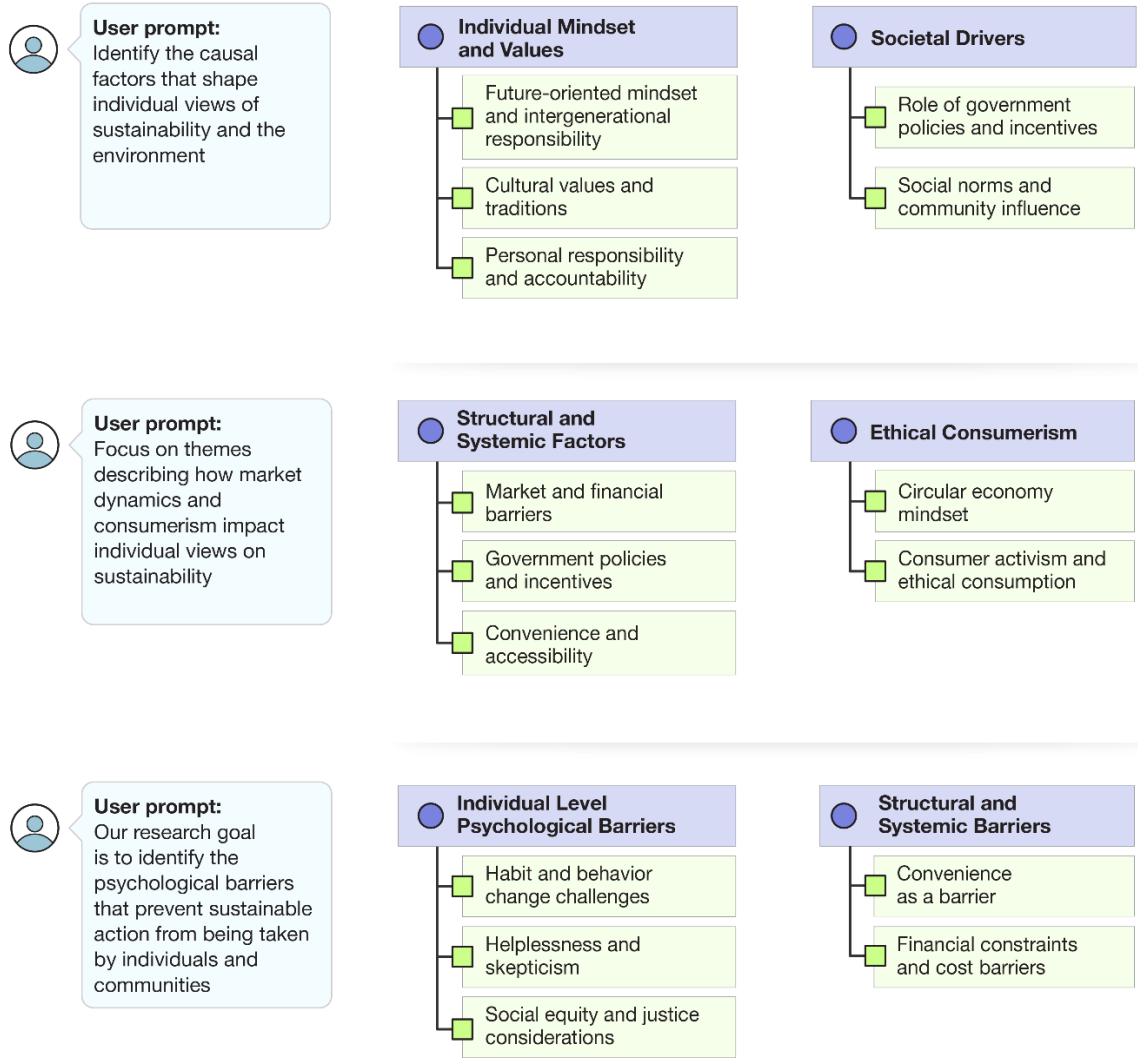

Figure 3: User prompts Muse to analyze the same dataset with different research lenses. As can be seen from the sample of generated parent/child codes, Muse identifies concepts that are more aligned to their researcher's specified analytical perspective while also identifying salient concepts that persist in the dataset independent of research perspective.

### 3.4 Code Application Methodology

Muse applies codes in two ways: *apply_code*, a feature which assigns a binary label or score to an excerpt given its relevance to a given code, and *distribute_code*, a feature which assigns one or multiple codes within a set of codes to an excerpt based on relevance.

Specifically, single-code application works by splitting documents into segments of approximately 80-words (modifiable by the user) and prompting an LLM to determine whether each is relevant to a given code (see next section for more details), with excerpts classified in batches to optimize both throughput and latency. The LLM is provided with the code name, definition, inclusion and exclusion criteria, and few-shot examples (when available).

Multi-code application (*distribute_code*) has a similar workflow; however, it differs in that it includes code information (definition, criteria, and few shot examples) for every code in the chosen subset. The user may specify if multiple codes can be assigned to a single excerpt (*allow multi-code*) and if every excerpt or document must be labeled with at least one code (*force-assign*). *apply_code* is designed for coding a single code at a time with high levels of user feedback and iteration, while *distribute_code* is designed for splitting parent codes into more granular subcodes, offering lower latencies than *apply_code* for this task but also less direct interaction with the researcher.

Examples of the core prompt texts leveraged by both *distribute_code* and *apply_code* are presented in Appendix A.3.

### 3.4.1 Inter-Rater Reliability as a Performance Metric

We assessed the performance of LLM annotations by measuring IRR with human coders across our evaluation dataset. While IRR is traditionally used to measure coding alignment between two human raters, here we use IRR as a metric for alignment between a human operator and an LLM annotator – with the degree of alignment reflecting how well the latter captures the opinions and preferences of the former [6].

The primary measure of IRR that we considered is Cohen's $\kappa$ [13], a standard metric in qualitative research that inherently adjusts for chance agreement between two coders and ranges in value from -1 (full disagreement) to 1 (full agreement). This is more holistic than percent agreement, or statistics like accuracy or precision, because it factors in the frequency of positives and negatives, which is often extremely unbalanced in qualitative datasets. Because the evaluation datasets vary greatly in size and number of codes, we sampled equally sized corpora from each of them such that the number of total words was similar.[1] To ensure robust statistical analysis while managing computational resources, we limited IRR evaluation to the top four most frequent codes per dataset (fewer when datasets contained fewer than four codes total, since codes with frequencies below 1% did not provide sufficient sample sizes for reliable evaluation,

### 3.4.2 Classification Strategy

LLMs produce outputs stochastically, and their responses can vary significantly with seemingly trivial input and output format changes - for example using a 4-point Likert scale versus a 6-point scale [25]. Given this sensitivity, we systematically explored two primary output formats for our annotation prompts to identify the most reliable approach. In *binarized scoring*, we prompted the LLM to produce a simple "Yes" or "No" response indicating whether a document excerpt related to a given code [33]. Alternatively, in *discretized scoring*, we prompted the LLM to rate each excerpt on a 1-10 relevance scale, allowing users to tune threshold confidence scores for determining final code assignments. When evaluating discretized scoring, we chose a threshold value that optimized Cohen's $\kappa$ in each code, simulating the performance a Muse user would observe if actively tuning this parameter within the platform. All subsequently reported Cohen's $\kappa$ values for discretized scoring in this work will leverage this convention.

### 3.4.3 Chain-of-Thought and Batch Size

Beyond output format, we examined several additional hyperparameters that influence coding consistency. Most notably, we tested the impact of chain-of-thought (CoT) prompting [49], which requests the LLM to provide brief explanations for its coding decisions prior to scoring, potentially improving both accuracy and interpretability [39]. We also systematically varied batch size (number of excerpts processed simultaneously) and the number of few-shot examples provided, as both factors can significantly affect model performance and computational efficiency.

---

[1] We sampled approximately 30,000 words of text from each dataset for hyperparameter optimization and 60,000 words for full evaluation.

### 3.5 Annotation Evaluation Results

Our evaluation of Muse annotations had three phases. First, we optimized the prompt design and output structure of the *apply_code* function through testing different combinations of hyperparameters and code classification approaches with one LLM (GPT-4o). Next, using the obtained optimal configuration, we benchmarked a range of closed-source and open-source LLMs to assess which performed the best and could be deployed at scale. Because the first phase of optimization focused on one LLM, there is an inherent bias in the hyperparameter selection that may favor this LLM over others. Future work should investigate the interactions between hyperparameter optimization and foundation model performance and explore frameworks for co-optimization of these components. Third, we compared performance of the *apply_code* function with the *distribute_code* function to identify for which types of datasets and codes the *distribute_code* function was preferred.

Table 3. Hyperparameter Optimization Grid Search (Sorted by Cohen's $\kappa$)

| Batch Size [a] | Few-Shot Examples | Scoring Type [b] | CoT Prompting [c] | F1 Score [95% CI] | Cohen's $\kappa$ [95% CI] |
|---|---|---|---|---|---|
| 5 | 4 | Discretized | Yes | **0.686 [0.611, 0.762]** | **0.545 [0.427, 0.662]** |
| 5 | 2 | Discretized | Yes | 0.670 [0.599, 0.742] | 0.532 [0.425, 0.639] |
| 10 | 2 | Discretized | Yes | 0.661 [0.577, 0.745] | 0.524 [0.408, 0.640] |
| 10 | 4 | Discretized | Yes | 0.663 [0.577, 0.748] | 0.521 [0.403, 0.640] |
| 5 | 2 | Binary | Yes | 0.657 [0.579, 0.734] | 0.518 [0.407, 0.628] |
| 10 | 2 | Discretized | No | 0.644 [0.563, 0.725] | 0.517 [0.406, 0.629] |
| 5 | 4 | Binary | Yes | 0.661 [0.583, 0.738] | 0.513 [0.398, 0.628] |
| 10 | 2 | Binary | Yes | 0.653 [0.574, 0.733] | 0.510 [0.399, 0.620] |
| 5 | 4 | Discretized | No | 0.652 [0.585, 0.719] | 0.508 [0.407, 0.610] |
| 5 | 2 | Binary | No | 0.645 [0.573, 0.717] | 0.508 [0.405, 0.611] |
| 10 | 4 | Binary | Yes | 0.652 [0.576, 0.728] | 0.507 [0.403, 0.610] |
| 5 | 0 | Binary | Yes | 0.651 [0.568, 0.735] | 0.502 [0.389, 0.616] |
| 5 | 0 | Discretized | Yes | 0.648 [0.566, 0.730] | 0.499 [0.387, 0.611] |
| 5 | 2 | Discretized | No | 0.634 [0.562, 0.705] | 0.498 [0.398, 0.599] |
| 10 | 0 | Discretized | Yes | 0.645 [0.566, 0.723] | 0.495 [0.385, 0.605] |
| 10 | 4 | Binary | No | 0.631 [0.556, 0.706] | 0.487 [0.381, 0.593] |
| 10 | 2 | Binary | No | 0.625 [0.550, 0.701] | 0.487 [0.379, 0.596] |
| 10 | 0 | Binary | Yes | 0.638 [0.563, 0.713] | 0.487 [0.384, 0.590] |
| 5 | 4 | Binary | No | 0.633 [0.560, 0.706] | 0.485 [0.382, 0.588] |
| 10 | 4 | Discretized | No | 0.623 [0.540, 0.706] | 0.481 [0.367, 0.596] |
| 10 | 0 | Discretized | No | 0.618 [0.535, 0.701] | 0.476 [0.366, 0.587] |

[a] Batch size is the number of excerpts included in a single prompt.
[b] Discretized scoring is a 1-10 score generated by the LLM (not log probabilities), whereas binary is a "yes" or "no" response.
[c] CoT prompting involves the LLM providing a short (1 sentence) explanation of its output prior to producing it.

*3.5.1 Hyperparameter Optimization Results*

The results of our hyperparameter tuning are displayed in Table 3. We observed that batch size did not significantly impact IRR while CoT improves performance for the GPT-4o model. We also found that discretized scoring, on average, produced higher IRR than its binary counterpart, with a confidence threshold of greater than or equal to 7 generally found to be optimal (see Appendix A.3 for a detailed view of confidence score thresholding). Providing few-shot examples also increased performance; however, we did not observe a clear difference between two positive examples versus four positive examples.

IRR varied significantly by dataset, code, and prompting strategy. More abstract topics that require an additional level of interpretation, and thus subjectivity, tended to have lower IRR between the LLM and human coder.[2] Such codes are often sentiment-related (e.g., "in favor," "against," or "neutral" with respect to Catalonian independence) or about the intent of language (e.g., "offensive language" or "hate speech"). These codes collectively averaged a Cohen's $\kappa$ of 0.23.

Another pattern that emerged was that codebook design considerations not directly considered by the *apply_code* function had a detrimental impact on IRR. Some codebooks require that at least one code be applied to each document and/or that no more than one code can be applied to a document; however, applying codes one label at a time does not provide this context or map cleanly onto such a problem set. For example, in the Catalonian independence dataset, researchers force-assigned one of the "in favor," "against," or "neutral" codes to every excerpt, a coding rule known to the research team but not currently communicable to Muse when applying codes one-by-one, resulting in low-performance. For such codes, the *distribute_code* (discussed below) function better aligns with human preferences because it can factor in mandatory selection between multiple codes and mutual exclusivity.

*3.5.2 IRR by Scoring Type*

One observation we made as we reviewed the best and worst-performing datasets was that there is an inherent level of sensitivity with which qualitative researchers approach a target concept. Even if a code is well-specified and contains inclusion and exclusion criteria, there is still often a level of human judgement as to whether it meets a certain threshold of salience that they desire for analysis. As detailed in the error analysis that follows, an LLM can require many iterations of human-curated samples to better align with a researcher's preferences and will not automatically classify themes with a researcher's desired level of sensitivity.

Binary scoring outputs prevent the end user from tuning this parameter after code application is complete. On the other hand, discretized outputs provide the user with the ability to tune LLM confidence thresholds as desired. Another approach to convert binary scoring into a higher-granularity score is through normalizing output log probabilities, a common post-processing approach for LLM-as-a-Judge systems [2, 48]. We assess the performance of log probabilities of "Yes" and "No" tokens in responses, which we map onto a 1-10 scale using a temperature-controlled softmax function [22], compared to other scoring types.

We present results both with a constant confidence score threshold optimized by model (untuned) as well as with simulating that a researcher has tuned their confidence score threshold for each code to align with their preferences (code-tuned). Figure 4 shows that the latter approach is optimal for achieving greater alignment and that generating discretized outputs - either by normalizing log probabilities of binary outputs or generating discretized scores directly - outperforms binary outputs. We chose to proceed with generating discretized scores as opposed to post-processing log probabilities because a) not all models output log probabilities, and output format varies by model, b) inaccurately extracting key answer tokens can lead to incorrect outputs [22], and c) while there could be further opportunities for post-processing or prompt optimization, our experiments show that discretized scoring outperforms binary log probabilities.

---

[2] Abstract or subjective topics are likely also more difficult for pairs of human judges to evaluate consistently [37].

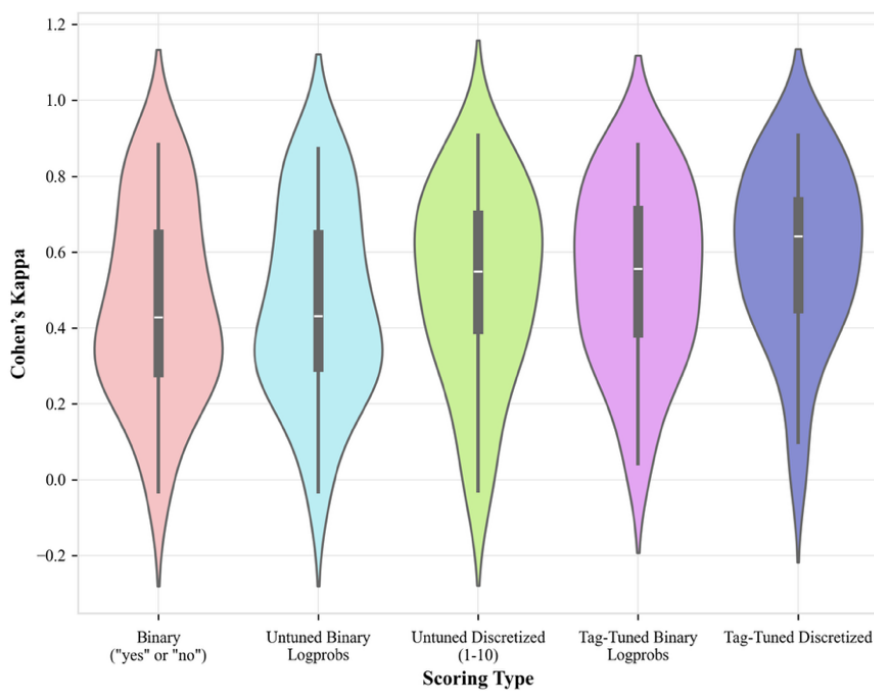

Figure 4: Distribution of IRR for each scoring type. The two best performing scoring types we evaluate are discretized (1-10) scoring and normalized log probabilities of binary "yes" or "no" responses, mapped onto a 1-10 scale. Code-tuned means setting the confidence score threshold for each code to align with the level of sensitivity in the human-coded data. Untuned means using a constant confidence score threshold for all datasets.

### 3.5.3 LLM Benchmarking Results

Once we arrived at optimal hyperparameters for GPT-4o (batch size of 4, discretized scoring, 4 few shot examples, and CoT prompting), we tested performance across a range of closed-source, open-source, reasoning, and non-reasoning LLMs with the full set of evaluation datasets.[3] Although we did not optimize prompts for each individual model, benchmarking IRR using one set of parameters provides a consistent and reproducible test arena for model comparisons. Additionally, it is of value to assess open-source model performance, as many real-world qualitative research environments may not be compatible with transmitting protected data to API providers. We tested a range of popular LLMs that were accessible to the team through Azure AI Foundry and Red Hat for cloud-based hosting with vLLM [28] but recognize that omissions remain that future work could evaluate (e.g., Gemini, Claude, and Qwen). Nonetheless, the results shown in Table 4 demonstrate a similar range of performance across state-of-the-art frontier models.[4]

Of the 15 models we tested, 8 outperformed GPT-4o by as much as 10-11% in IRR with both untuned and code-tuned confidence score thresholding. Most notably, GPT-4.1 exhibited the highest level of performance - demonstrating parity with newer foundation models like o3 and GPT-5 - although this may be a relic of the fact that the performed GPT-4o hyperparameter optimization maps most directly onto this model. Using an unweighted ensemble of the discretized score of three of the top frontier models (GPT-4.1, o3, and Grok 3) achieves an untuned $\kappa$ of 0.55, code-tuned $\kappa$ of 0.61, and for codes with a clear definition (n=11), a $\kappa$ of 0.71.[5]

---

[3] We use a batch size of 4 instead of 5 (optimal value in hyperparameter optimization) because some smaller models tested failed to consistently produce correctly formatted structured outputs for more than 4 documents at once. There is no significant performance difference between the two batch sizes.
[4] The different in absolute score for GPT-4o in Table 3 and Table 4 is due to hyperparameter optimization including only a subset of datasets and a smaller sample of total codes (21 codes in Table 3 as opposed to 33 codes in Table 4). To test for robustness in IRR with all datasets, we compared untuned Cohen's $\kappa$ for GPT-4.1 with a range of sample sizes from each dataset (50,000 to 200,000 words, default = 60,000) and number of codes per dataset (top 2 to top 10, default = top 4) and found that average untuned Cohen's $\kappa$ ranged from 0.44–0.52 and code-tuned Cohen's $\kappa$ ranged from 0.52–0.59.

Table 4. Inter-Rater Reliability by LLM

| Model Name | Optimal Confidence Threshold (≥)[a] | Untuned Cohen's $\kappa$ [95% CI] | Code-tuned Cohen's $\kappa$ [95% CI][b] |
|---|---|---|---|
| gpt-4.1-2025-04-14-global | 8 | 0.515 [0.426, 0.604] | **0.593 [0.512, 0.674]** |
| o3-2025-04-16-global | 9 | **0.520 [0.423, 0.617]** | 0.590 [0.508, 0.672] |
| gpt-5-2025-08-07-us (default) | 9 | 0.515 [0.422, 0.608] | 0.578 [0.498, 0.658] |
| grok-3-v1 | 8 | 0.505 [0.425, 0.585] | 0.572 [0.496, 0.648] |
| grok-3-mini-v1 | 9 | 0.492 [0.402, 0.582] | 0.553 [0.470, 0.636] |
| o4-mini-2025-04-16-global | 9 | 0.492 [0.398, 0.586] | 0.546 [0.459, 0.633] |
| gpt-oss-120b (high reasoning) | 9 | 0.486 [0.386, 0.586] | 0.543 [0.450, 0.636] |
| Microsoft MAI-DS-R1 | 9 | 0.499 [0.413, 0.585] | 0.539 [0.456, 0.622] |
| gpt-4o-2024-08-06-us | 8 | 0.461 [0.372, 0.550] | 0.536 [0.453, 0.619] |
| gpt-oss-120b (low reasoning) | 8 | 0.483 [0.388, 0.578] | 0.535 [0.446, 0.624] |
| Llama-3.1-Nemotron-Ultra-253B-v1 (reasoning) | 9 | 0.470 [0.382, 0.558] | 0.525 [0.440, 0.610] |
| DeepSeek-V3-0324 | 8 | 0.464 [0.381, 0.547] | 0.525 [0.445, 0.605] |
| DeepSeek-R1-0528 | 9 | 0.471 [0.384, 0.558] | 0.524 [0.445, 0.603] |
| Llama-3.3-Nemotron-Super-49B-v1 (reasoning) | 9 | 0.466 [0.376, 0.556] | 0.516 [0.434, 0.598] |
| Llama-3.1-Nemotron-Ultra-253B-v1 (non-reasoning) | 9 | 0.428 [0.344, 0.512] | 0.488 [0.406, 0.570] |
| Llama-3.3-Nemotron-Super-49B-v1 (non-reasoning) | 9 | 0.429 [0.343, 0.515] | 0.484 [0.403, 0.565] |
| gpt-4.1-mini-2025-04-14-global | 9 | 0.383 [0.299, 0.467] | 0.454 [0.372, 0.536] |
| Nemotron-H-47B-Reasoning-128K (reasoning) | 10 | 0.379 [0.296, 0.462] | 0.420 [0.343, 0.497] |
| Nemotron-H-47B-Reasoning-128K (non-reasoning) | 10 | 0.290 [0.216, 0.364] | 0.332 [0.262, 0.402] |

[a] Optimal confidence threshold is a constant (untuned) threshold for assigning positive and negative codes based on 1-10 scores generated by the LLM that optimizes IRR for a given model. Reasoning models (e.g., o3 and Grok 3 Mini) tend to have a higher optimal confidence score threshold than non-reasoning models.
[b] Code-tuned means setting a separate confidence score threshold for each code to align with the level of sensitivity in the human-coded data.

Distilled, non-reasoning models, like gpt-4.1-mini and the Nemotron 49B and 47B models, lag behind full-sized and/or reasoning-based models considerably, while distilled *reasoning* models like o4-mini and Grok 3 Mini are much closer in IRR to top-performing models. OpenAI's gpt-oss-120b and Microsoft's post-trained version of Deepseek-R1 (which fills information gaps in the original model) are the most aligned open-source models. Because gpt-oss is significantly smaller (nearly 6 times fewer parameters), it is the preferred open-source choice for optimizing cost, latency, and throughput.

---

[5] Well-specified codes include the four arXiv abstract codes with specific definitions (Cohen's $\kappa$ range of 0.69-0.87), the four eating disorder-related codes in the Reddit data set (Cohen's $\kappa$ range of 0.73-0.91), two of four Wikipedia discussion codes ($\kappa$ = 0.65 and 0.45), hate speech defined in the hate speech detection dataset ($\kappa$ = 0.46), and trauma defined in the genocide tribunal dataset ($\kappa$ = 0.55). Catalonian Independence dataset excluded because without additional criteria, it isn't obvious what having a favorable or negative view toward Catalonian Independence means. Other codes with short definitions are excluded because of a lack of clarity with respect to their target concept (e.g., "informing seeking").

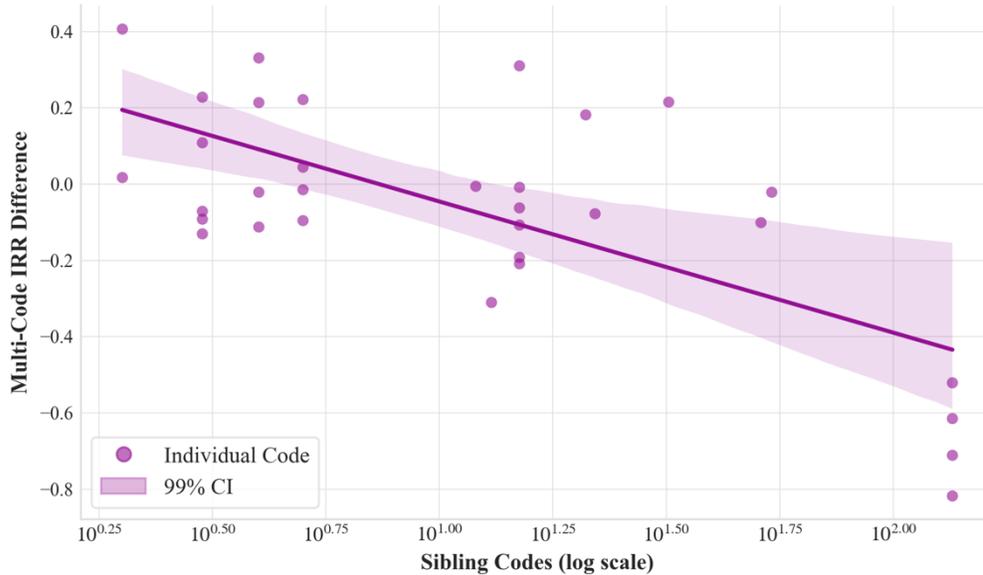

Figure 5: Difference in IRR (Cohen's $\kappa$) between multi-code application and single-code application with respect to the number of sibling codes. A greater number of sibling codes (i.e., options to choose from for multi-code application) is associated with a decrease in IRR compared to single-code application. While there is a wide range of variation, *distribute_code* outperforms *apply_code* on average when there are approximately 6 or fewer sibling codes and performs worse than *apply_code* when there are 10 or greater sibling codes.

*3.5.4 Multi-Code Application*

To assess the performance of the *distribute_code* function, which applies one or multiple codes within a set of codes to an excerpt based on relevance, we compared its IRR with the *apply_code* function. For datasets with a hierarchical codebook, we provided all siblings of a target code as options, while for flat codebooks, we provided all codes as options. Figure 5 shows that much of the variation ($r^2$=0.46) in the difference in IRR between the two functions is a function of the number of sibling codes considered by the *distribute_code* function. When the number of alternative options is low (less than about 6), *distribute_code* outperforms the *apply_code* function and results in lower latency and cost as it requires fewer prompts. When a codebook contains constraints such as *force-assign* or no *multi-code*, the *distribute_code* function is strongly preferred and yields IRR scores that are significantly higher (Cohen's $\kappa$ greater by an average of 0.18 for such problematic codes under *apply_code*).

## 4 ERROR ANALYSIS

While Muse applies annotations at an IRR comparable to humans, inevitably, not all annotations match those of qualitative researchers. We conduct an error analysis of inconsistencies between human and Muse codings to categorize failure modes and identify areas for future improvement.

### 4.1 Error Codebook Design

To gain a better understanding of the types of errors made by the LLM and inconsistencies with human coders, our team analyzed a random sample of errors. For tractability, we narrowed our focus to two datasets, Topic Annotations on Reddit

Posts, which contained label definitions, more document context, and overall clearer organization than many other datasets in our repository, and the ATLAS.ti Sustainability dataset, a sample interview-based project. Because the Reddit dataset consists of social media posts rather than technical literature, and the Sustainability dataset is a sample project intended to be understood by researchers in any area, the language and content are more accessible for comprehensive error analysis compared to specialized domains requiring deep subject-matter expertise, such as medical documents or scientific abstracts.

We ran the *apply_code* function across the top eight most frequent codes in the Reddit dataset and the top four most frequent codes in the Sustainability dataset using the average score of an equally weighted ensemble of GPT-4.1, o3, and gpt-oss-120b models to not overfit our error analysis to any one foundation model.[6] Under these conditions, the Cohen's $\kappa$ score for the top eight codes in the Reddit dataset averaged 0.784, and averaged 0.321. We sampled 50 false positives (FPs) and 50 false negatives (FNs)[7] from each and analyzed them qualitatively by considering

- Ground-truth positives associated with each target code
- The complete set of codes applied by human researcher to a considered excerpt
- Target code label and definition
- Reasoning strings produced by the LLM
- LLM confidence scores

Table 5. Muse Error Codebook

| Code | Definition |
| --- | --- |
| Code under-specification | The target code is defined too broadly and/or has ambiguity in its application criteria. This includes codebooks where multiple codes may have overlapping scope and application criteria. |
| Contextual limitation | Insufficient surrounding text or document context in excerpt to assess applicability. This includes tangential mentions of the target theme that are not salient concepts in the document. |
| Annotation inconsistency | Application of code is a deviation from its definition, criteria, and other positive examples. |
| Thresholding error | Assigned confidence score and reasoning are aligned with human code, but confidence score threshold is suboptimal for the particular excerpt. |
| Other LLM error | Other inexplicable error by the LLM. Reasoning and confidence score are incorrect. |

After an initial review by two authors, we iteratively developed and refined the codebook in Table 5 to describe the root causes of the observed errors. Two other authors, who are experienced qualitative researchers with backgrounds in anthropology and health policy, independently coded the set of errors for the presence of each code in the codebook and then jointly resolved discrepancies. The first round of coding resulted in an average percent agreement of 68.2% and average Cohen's $\kappa$ of 0.11 across codes and datasets. After several meetings between the coders to resolve differences in interpretation of codes and reconciliation of inconsistencies between them (e.g., human coder vs. LLM error and breadth of application of codes), they reached an average percent agreement of 96.9% and Cohen's $\kappa$ of 0.91.

---

[6] We used only the top four most frequent codes in the Sustainability dataset because the frequency of each code was much lower than in the Reddit dataset, limiting the number of examples in the test after holding out few-shot examples.
[7] Because of the smaller size of the Sustainability dataset and also smaller number of total codes, it resulted in only three total FNs. We reviewed all three but recognize this is a small sample.

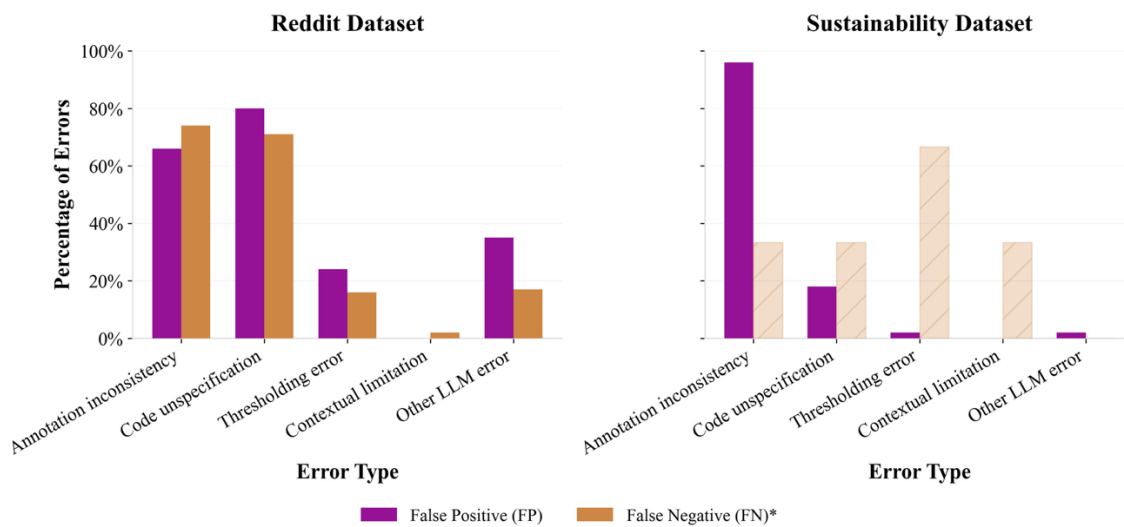

Figure 6: FPs and FNs by error type and dataset. Percentage of errors is an average frequency of the codings of both coders. The two most common error types for both FPs and FNs across datasets are code under-specification and annotation inconsistencies (deviations from the establishing coding scheme). *FNs for the Sustainability dataset are made transparent because the dataset only had three FN examples to code, so the percentage of errors is not representative.

## 4.2 Error Analysis Results

As shown in Figure 6, a large proportion of errors made by the LLM (66% of Reddit FPs, 74% of Reddit FNs, and 96% of Sustainability FPs) are explained by annotation inconsistencies where the human-assigned code deviated from the stated definition of a code. This could potentially be due to coder fatigue or interpretive drift, and/or in the case of the Sustainability dataset, overly complex and long codebooks leading to inconsistent application of sub-codes. The observed inconsistencies highlight the broader challenge of establishing reliable ground truth in qualitative research and ultimately may reduce our ability to definitively assess LLM performance against human coders.

While the Reddit dataset had a better specified codebook than many of the other evaluation datasets, most errors (80% of FPs and 71% of FNs, 18% of Sustainability FPs) were also incorrectly labeled in part or in full due to under-specified target codes lacking specific definitions. This often leads to code confusion between semantically similar codes with insufficient disambiguation criteria. The most frequent example of such confusion in the sample was between the code "thinspiration," defined as "drive for thinness, want to be thinner or skinny," and the code "weight loss," defined as "body weight loss."

In addition to suboptimal confidence score thresholding (24% of Reddit FPs, 16% of Reddit FNs, 2% of Sustainability FPs), some remaining LLM responses were inexplicably erroneous (35% of Reddit FPs, 17% of Reddit FNs, 2% of Sustainability FPs). These errors show that optimizing confidence score thresholding, using an ensemble of high-performance models, and clearly specifying codes may still result in some unaligned outputs.

A common characteristic of FPs that we observed in the Reddit dataset was a high density of additional ground truth codes other than the FP target code. 31 of 50 FPs we reviewed contained at least three alternative positive codes, significantly above average for this dataset. We posit that the presence of many alternative codes influences the likelihood of a FPs through both diminishing the relative salience of each individual code in comparison to alternative options as well as by potentially inducing greater levels of coder fatigue.[8] While we do not know definitively at which point or whether the human annotators were fatigued, the requirement of assigning a high number of codes to a single passage can be seen as somewhat of a proxy for requiring high attention to detail and resulting fatigue.

It follows that one logical extension of Muse is as an auditing tool for qualitative research teams to assess the consistency of coding across coders and throughout a dataset to check and correct for coder drift and fatigue.[9] We simulate such an audit in Figure 7, which displays a rolling average of the FP and FN rates over each chronological documents in the Reddit dataset (likely over time). Both rates experience a modest increase over the final 200 documents in the dataset, indicating that the coders may have become fatigued toward the end of manual coding. Assuming that documents are not clustered in terms of difficulty, ideally a human research team would re-evaluate codes and spend more time with coders to work toward codings that do not have a drift in performance over time.

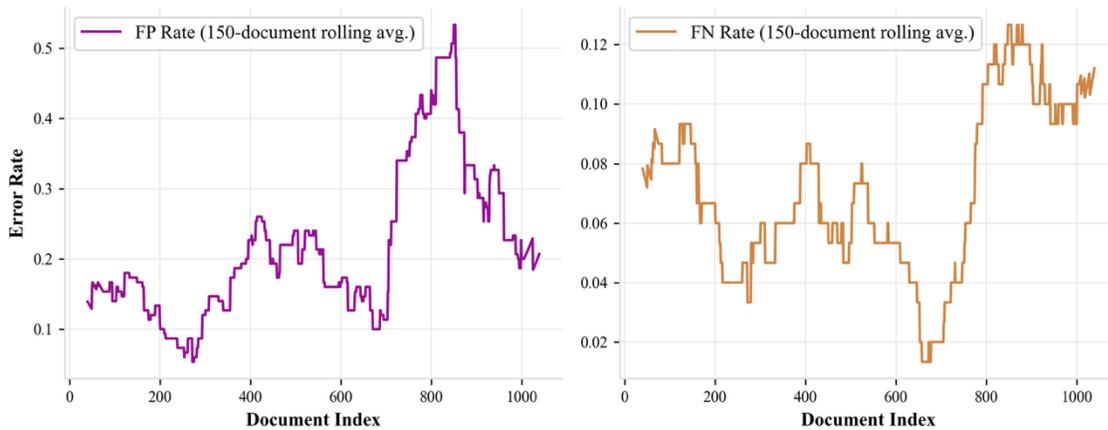

Figure 7: Rolling average of FP and FN rate by document index (over time of coding). FPs and FNs increase modestly over the final 20% of documents, suggesting that coder(s) may have experienced fatigue or drift toward the end of manual coding.

Given this error analysis, if the following easily implementable steps are taken, we believe the error rate can be reduced substantially. First, codes should be fully specified. This means providing complete concept definitions that provide clarity about what specifically is of interest to the researcher and what should be included and excluded. Second, if two codes are semantically similar, the researcher should describe how they differ and when one should be applied or both should be applied. If both appear to be describing the same underlying construct, the researcher should consider merging them. Finally, from the perspective of system development, some errors due to a lack of full document context could be addressed by processing full documents in single prompts as opposed to chunked excerpts - a more complex task yet one increasingly in reach as LLM context windows grow.

---

[8] High code density could also be a function of very rich text that touches on multiple domains. However, this presumably still would require a greater level of attention to detail to comprehensively annotate.

[9] Some qualitative researchers likely view interpretive drift and coder fatigue as natural, or even valuable, parts of the qualitative research process that could be a sign of evolving understanding of complex data through reflexivity and refinement. However, emergent changes in interpretation require multiple rounds of recoding [42] and could threaten interrater reliability between multiple coders and extend project timelines due to reconciliation of inconsistencies.

### 4.3 Using FPs and FNs as Few-Shot Examples

Muse is guided by human-provided few-shot examples, which serve as visceral representations of a given code's definition and inclusion/exclusion criteria. However, at least amongst the datasets we explored, codebooks in qualitative research are often under-specified and require hours of discussion and instruction with coders to articulate. To address this challenge, we tested whether providing inconsistencies between Muse and human codings back into Muse as additional few-shot examples would improve alignment on remaining documents. This mode of iterative computer-human interaction is a core feature of the platform as users can correct unaligned LLM codes and revise their codebooks after code application.

To simulate a live-user interacting with the platform, we ran the *apply_code* function on a given code and fed the false positives and false negatives back into Muse through few-shot examples before running the feature again, performing this iterative cycle up to six times per code (shown in Figure 8).

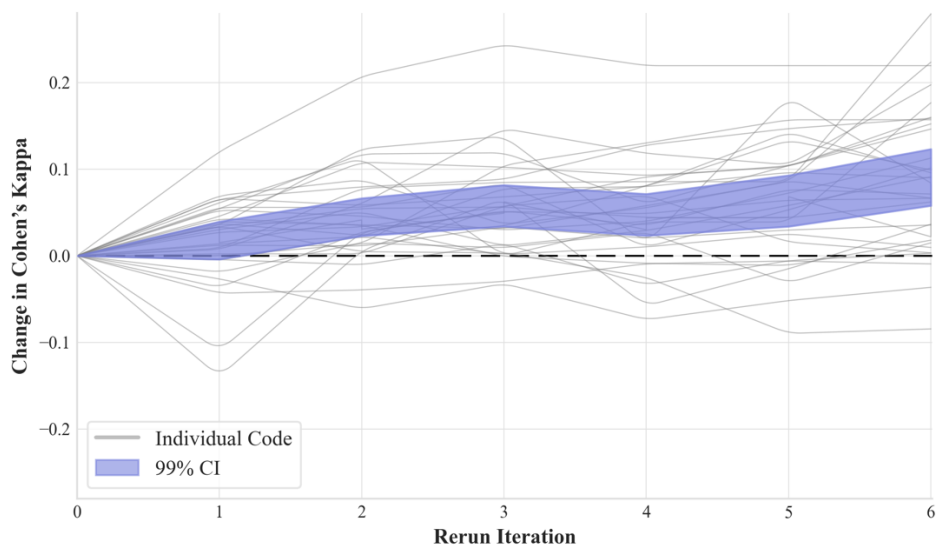

Figure 8: Change in IRR after providing errors as few-shot examples. Iteratively providing a small number of errors as examples increases IRR above baseline through at least six iterations. Baseline (Rerun iteration 0) uses four random examples. Each rerun iteration adds an additional four examples (2 FPs and 2 FNs). IRR is calculated for each iteration using the remaining subset of documents not used as examples after rerun iteration 6.

On average across all datasets, we found that providing four errors (2 FPs and 2 FNs) each iteration as additional examples resulted in a non-trivial increase in Cohen's $\kappa$ (+0.05 after 3 iterations and +0.09 after 6 iterations).[10] Codes that experienced the greatest increase in IRR from iterative code application were those that contained vague definitions such as "*acq*: mergers/acquisitions" in the Reuters dataset and "*privacy*: restricted access" in the interview dataset on data curation. Another type of typing of coding error that was addressed by this iterative approach was codes with similar meaning in a codebook that could be confused with one other without sufficient inclusion and exclusion criteria. For example, the arXiv dataset's "High Energy Physics – Phenomenology" and "High Energy Physics – Theory" codes increased in Cohen's $\kappa$ by 0.21 and 0.12 respectively after just two iterations as ambiguities between them were resolved through FP and FN examples.

---

[10] Baseline had four random examples. By the sixth iteration, the total number of examples was 28 (four random, 12 FPs, and 12 FNs).

On the other hand, codes that required additional context beyond what was provided in the document or had codebook constraints such as code mutual exclusivity did not experience a performance increase. For such cases, FP and FN examples do not capture the reason for the coding error. In many of these cases, there are hidden coding rules the researcher is aware of that the LLM cannot access (i.e. *multi-code, force-assign, etc*.), and thus providing such examples does not reduce ambiguity. Overall, this experiment demonstrates the value in human engagement with LLM outputs and the capacity to significantly improve alignment through human feedback.

## 5 LIMITATIONS

Our evaluation has three main limitations: dataset quality, the inherent subjectivity of qualitative research, and limited dataset diversity.

To the first point, dataset quality issues significantly impact our benchmarking accuracy and are driven by multiple factors. For example, we lack context about the original research teams' analytical intentions, cannot account for potential coder fatigue in ground truth annotations, and benchmarked against many datasets that lacked clear code definitions and inclusion/exclusion criteria. These factors introduce noise that likely caused misidentification of FPs and FNs in our benchmarking.

Consequently, our reported $\kappa$ values contain some inaccuracy, although based on our error analysis, these reported values likely represent lower bounds rather than upper bounds of true performance. While this noise does not substantially affect our hyperparameter optimization - which relies on relative improvements in $\kappa$ rather than absolute values - it does limit our ability to assess Muse's absolute alignment with human analytical practices. This gap between benchmarked performance and real-world alignment remains a critical factor in determining whether AI-assisted systems like Muse can meet the standards required for rigorous qualitative research.

Similarly, our optimization process is inherently biased toward the specific analytical perspectives embedded in our evaluation datasets. Qualitative research embraces interpretive diversity - the same dataset can legitimately yield multiple valid codebooks depending on theoretical framework, research questions, and analytical goals. By optimizing against ground truth data produced by a single research team, our system learns to replicate particular researchers' interpretive choices and biases rather than developing a generalizable approach to qualitative analysis.

Dataset diversity was another constraining factor, in that certain types of qualitative data, such as focus groups or ethnographic observations, were underrepresented in the dataset. Thus these findings may not be generalizable to the full spectrum of qualitative research. Similarly, our datasets were primarily written in, or translated to, English – meaning the extent to which these methods can be extended to other languages remains untested.

Our analysis also focused on the most frequent codes in our dataset to ensure sufficient sample sizes for our metric calculations. However qualitative research often draws on rare but theoretically important themes. Our evaluation provides no insights into AI performance on these less frequent codes.

Finally, our evaluation focuses primarily on static benchmarking against previously completed research projects, which cannot capture the dynamic, iterative nature of real qualitative research practice. The central value proposition of Muse lies in interactive collaboration between researchers and AI systems - enabling iterative exploration of research frameworks, real-time codebook refinement, and adaptive analytical approaches. Our current evaluation methodology, while necessary for establishing baseline performance, assumes fixed research questions and static codebooks that do not reflect how qualitative researchers engage with their data.

This limitation points to a critical need for user studies that can assess Muse's effectiveness in authentic research contexts. Future evaluations should examine how human-AI collaboration affects research outcomes through controlled experiments comparing researchers working with and without AI assistance, measuring not only efficiency and coding consistency but also analytical creativity, theoretical development, and overall research quality. Such studies will be essential for understanding whether AI assistance enhances or constrains the interpretive processes that define rigorous qualitative inquiry.

## 6 FUTURE OUTLOOK

Building on the error analysis discussed earlier, we have identified several promising enhancements to the Muse system that we plan to explore in future work. These improvements are designed to increase both usability and effectiveness, with implications for broader human-AI collaboration in qualitative research.

*Enhanced Contextual Coding*: Within the *apply_code* and *distribute_code* feature, we will provide the LLM with access to entire documents rather than isolated text snippets during annotation. This expanded context should reduce misapplied codes caused by insufficient understanding of document themes and prevent neglect of more subtle but important concepts. Implementation will likely require transitioning from LLM classification to span annotation techniques, which have shown promise in recent work [24].

*Researcher-Guided Coding Rules*: We plan to elicit more detailed information about researcher' coding frameworks and analytical intentions. Specifically, we will enable researchers to specify mutually exclusive code relationships (preventing simultaneous application of incompatible codes) and implement automated protocols for detecting highly correlated codes that may warrant merging. These features should significantly reduce the "code confusion" and contextual errors identified in our analysis.

*Real-World Evaluation and Human-AI Collaboration Studies*: Moving beyond benchmarking, we will conduct user studies and controlled experiments to assess the practical utility of the Muse platform and its impact on qualitative research processes. Key research questions include understanding how human-AI collaboration affects research output quality, the diversity of generated codebooks, and researchers' overall experience with qualitative analysis.

*Addressing Bias and Maintaining Research Rigor*: We are particularly interested in examining how human-AI teaming might inadvertently enable individual researchers to introduce unchecked biases - a concern that becomes more pressing when AI reduces the need for larger research teams that traditionally serve as checks against such bias. We aim to explore workflows that leverage AI capabilities while preserving collaborative team environments that promote diverse perspectives and analytical rigor.

## 7 CONCLUSION

The integration of LLMs into qualitative research workflows represents both an opportunity and a challenge for the field. Through our comprehensive benchmarking of the Muse platform across eleven diverse datasets, we demonstrate that AI-assisted qualitative analysis can achieve human-level inter-rater reliability (Cohen's $\kappa \approx 0.7$) while offering capabilities that extend beyond traditional computational approaches. Our systematic evaluation reveals that LLM-based systems can consistently apply established coding schemes, generate empirically grounded codebooks, respond to researcher guidance in ways that many other computational approaches cannot, and serve as a check on human bias.

However, our findings also underscore important limitations. The inherent subjectivity of qualitative coding means that no single algorithmic approach can capture the full range of valid analytical perspectives. Our error analysis reveals that even well-optimized systems struggle with under-specified codes, semantic ambiguity between related concepts, and the contextual nuances that human researchers navigate intuitively. These challenges remind us that AI assistance should augment rather than replace human interpretive expertise.

Looking forward, the value of AI-assisted qualitative research lies not in automating analysis but in enabling researchers to work at unprecedented scales while maintaining methodological rigor. By providing tools that can explore multiple theoretical frameworks simultaneously, maintain coding consistency across large corpora, and surface patterns that might escape human attention, systems like Muse open new possibilities for qualitative inquiry. The key to realizing this potential will be developing human-AI collaboration workflows that leverage computational efficiency while preserving the interpretive depth, theoretical sophistication, and contextual sensitivity that define rigorous qualitative research. As these tools mature, the qualitative research community must continue to critically evaluate their impact on research practices, ensuring that efficiency gains do not come at the cost of analytical quality or methodological integrity.

## A APPENDICES

### A.1 Dataset Search and Preprocessing

*A.1.1 Search Strategy*

We identified the text classification datasets by reviewing benchmark datasets used in previous machine learning research on text classification [42, 51]. To locate additional datasets from qualitative research projects, we searched Syracuse University's Qualitative Data Repository, UK Data Service, Center for Open Science, Papers With Code, TU Darmstadt's TUdatalib, Kaggle, and Google Dataset Search for annotated qualitative data and interviews. The search was not fully exhaustive of each database and was most time-intensive with respect to finding interview-based datasets, which are rarely released along with research projects due to data protections. The two interview-based datasets we identified were minimally redacted (de-identification and two omitted transcripts) and contained full author codings. We chose to terminate the search once we found a satisfactory level of diversity in project domain, data format, and codebook complexity. Although we reviewed over 30 candidate datasets, we chose not to include additional datasets in domains that were already covered by the eleven selected datasets (e.g., additional news headlines), did not contain full interview transcripts, or lacked a consolidated file with annotations (for example, individual transcripts with highlight markings and margin notes).

*A.1.2 Quality Score Rubric*

We implemented a rough 1-10 dataset quality score through an unweighted average of scores we assigned for definition completeness, annotation consistency, and document completeness.

A score between 1-3 for document completeness means that few, if any, definitions are provided, and label names are not specific enough to apply the code. A score between 4-7 means that most labels contain a definition or descriptive name, but definitions are generally too short or vague to be of much use. A score of 8 or higher means that specific definitions containing criteria and nuanced theme understanding are provided.

For annotation consistency, a score of 1-3 means that codes are applied without any regularity or experience significant inconsistencies because of study design or execution. A score of 4-7 means that code application generally aligns with stated criteria but has some deviations. A score of 8 or higher means no noticeable deviations in code application are present.

Document completeness is a measure of whether is sufficient context provided in documents to replicate the author's codings. A score of 1-3 means that is impossible to replicate codings because the author(s) referenced other material to make their judgements that is not provided. A score of 4-7 means that some document context is omitted for each code, but codings are still made at the level of each document and can mostly be replicated. A score of 8 or higher means that very few document context concerns exist.

*A.1.3 Data Preprocessing*

For consistent comparison of human annotations with Muse across the datasets, we attempted to standardize the evaluation datasets and their respective codebooks. This process involved converting datasets to comma-separated values format from QDPX, XMI, and other file types; extracting majority-vote labels from the two datasets with multiple annotators (Wikipedia and Hate Speech); and creating codebooks for datasets lacking them including adding label definitions, inclusion/exclusion criteria, and code hierarchy when available within each dataset's associated references.

For interview data, Muse only assigns codes to individual responses to interview questions to avoid coding statements from interviewers rather than interviewees. However, not all evaluation datasets followed this convention, with code annotations often spanning multiple conversation turns and often varying substantially in length. To reconcile this difference, we assigned "ground truth" positive codes to the question response closest to each human-coded annotation in terms of maximizing word overlap.[11] For the remaining datasets, which had documents of much shorter length than interviews, we performed document-level aggregation: we collected all codes assigned by Muse to any text segment within each document, collected all codes assigned by human annotators to any text segment within the same document, and then compared these two document-level code sets to evaluate assignment accuracy.

## A.2   Codebook Similarity Score

For every pairwise combination of codes between LLM-generated ($C_L$) and human-generated ($C_H$) codebooks, we computed:

1. **Semantic similarity ($S_{sim}$)**: Cosine similarity between sentence embeddings of code labels and descriptions
2. **Structural similarity ($S_{str}$)**: Hierarchical position, path length, and subtree characteristics within each codebook where $S_{str}(c_1, c_2) = s_{level}(c_1, c_2) + s_{path}(c_1, c_2) + s_{subtree}(c_1, c_2)$ where $s_{level}$ measures depth similarity, $s_{path}$ compares root-to-node distances, and $s_{subtree}$ evaluates organizational complexity beneath each code.

These components were combined into a composite similarity score, then optimally aligned using the Hungarian algorithm to find the best global matching between codebooks. After matching all codes from $C_H$ to their associated codes in $C_L$, we calculated the average weighted similarity score across each pair, $S_A = \alpha S_{sim} + \beta S_{str}$, providing a final codebook similarity score, $\overline{S}_A$, averaged across all pairs that considers both content and organizational logic. Here, $\alpha + \beta = 1$, and both variables were varied across the interval [0.1, 0.9], with the final $\overline{S}_A$, value averaged across these weightings to ensure balance across both similarity components.

---

[11] For the Data for Search Systems Study, annotations did not fully correspond to the original text in interview transcripts. To reconcile differences such as missing words or summary of some responses, we used GPT-4.1 mini to first return the original excerpt in an interview transcript corresponding with each annotation before matching it to the closest 80-word chunk.

## A.3 Confidence Score Thresholding

Using discretized LLM scoring (e.g., 1-10) requires selecting a score threshold for discriminating between positive and negative cases effectively. For hyperparameter optimization with GPT-4o, a threshold for scores seven or high was optimal for achieving the highest IRR. For the full evaluation test with all LLMs, thresholds of ≥8 or ≥9 were optimal. Because researchers have varying levels of desired sensitivity to concepts of interest, selecting a confidence score threshold for each individual code (code-tuned) increases IRR. Figure 9 displays the tradeoffs in calibrating a confidence score threshold in terms of sensitivity versus specificity during the hyperparameter optimization run with GPT-4o. As a confidence score threshold increases, the FP rate decreases, but the FN rate increases.

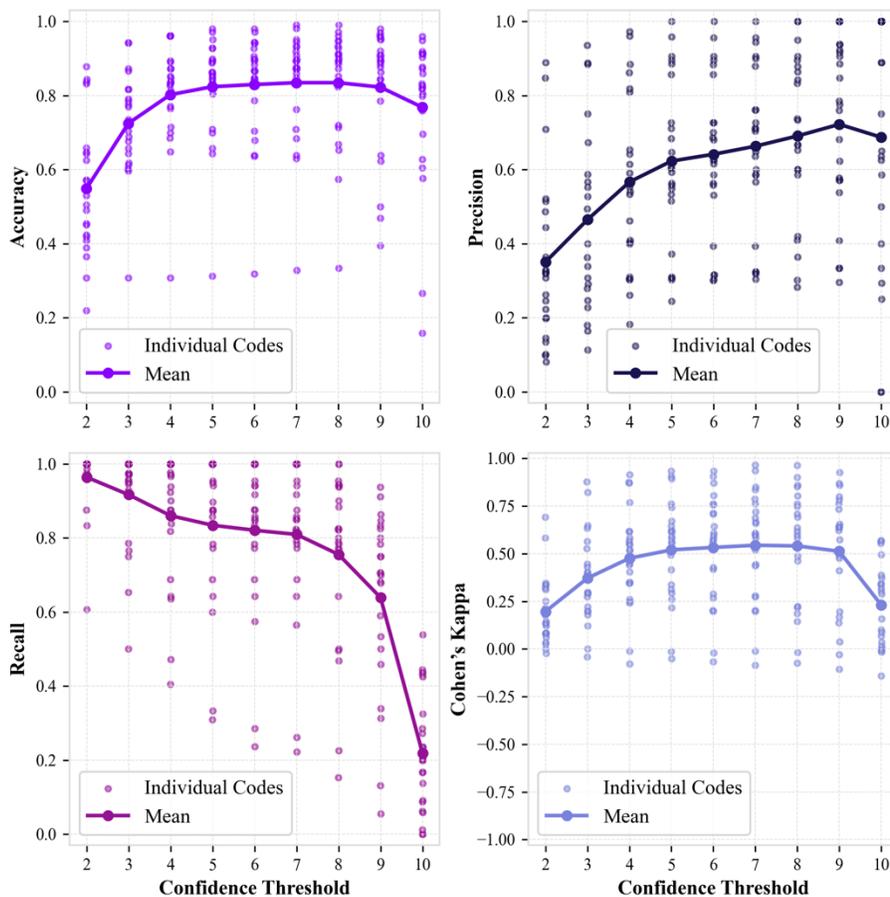

Figure 9: Performance metrics (accuracy, precision, recall, and Cohen's $\kappa$) by confidence score thresholds 2-10. A confidence score threshold $i$ means that only scores assigned by GPT-4o greater than or equal to $i$ count as positives.

## A.4 Annotation Prompts

*A.4.1 Single-Code Application*

The prompt text leveraged for the *apply_code* function is given below:

```
# Qualitative Research Task

You are assisting a qualitative researcher with tagging interview statements. Your task is to decide
whether each provided statement should be tagged with the code: **`{tag}`**.

You will use the provided **code definition**, **tagged examples**, **untagged examples**, **inclusion
criteria**, and **exclusion criteria** to make your decision. Your decision-making process should
prioritize:

1) Alignment with the code name/definition.
2) Satisfying at least one inclusion criterion while satisfying none of the exclusion criteria.
3) Consistency with provided positively tagged examples, avoiding similarities to negatively tagged
examples.

A worked-out example using the **AI for literature review tag** is provided to guide your decision-
making   process.

---

### Output Instructions

For each statement, provide:
1. A brief explanation (5-10 words) of your reasoning
2. A confidence score (1-10) where 10 indicates highest confidence the statement should be tagged and
1 indicates the statement should definitely not be tagged

Your output will be a JSON with one numeric key for every provided input statement. The value for
each key will be a JSON with two keys: "REASONING" and "SCORE".

---

### Worked Example: Literature Review Tag

**Code Name**: Using AI for literature reviews

**Code Definition**:
This code applies to statements that reference the use of Artificial Intelligence (AI) or Machine
Learning (ML) to assist with literature reviews. Specifically, it focuses on how AI is applied to
discover, organize, or summarize academic or scientific publications.

**Inclusion Criteria**:
1. Mentions AI or ML in the context of finding, summarizing, or organizing academic articles or
studies.
2. Names a specific AI model or approach (e.g., GPT, BERT) used to conduct or facilitate a literature
review.
3. Clearly states "literature review" (or a recognizable synonym, such as "systematic review") while
also describing an AI-based process.
```

**Exclusion Criteria**

1. Mentions the word "plagiarism" (indicating AI is being used for plagiarism detection instead of literature review).

2. Describes AI usage for purposes unrelated to literature reviews (e.g., AI for financial forecasting, image classification).

3. Focuses on manual processes or workflows unrelated to AI (e.g., manually reviewing case studies without any AI involvement).

**Tagged Examples (Relevant Statements)**:

1. "We developed a GPT-based tool to summarize new articles for our scoping review."

2. "Our research uses AI-based clustering to find relevant papers, automating the literature review process."

**Untagged Examples (Irrelevant Statements)**:

1. "An AI model helped detect plagiarism in graduate theses before publication."

2. "Our study uses advanced ML to analyze survey data from an ongoing clinical trial."

#### Statements to Evaluate

1. "During our systematic review, we employed both GPT-4 for synthesizing findings and traditional manual screening methods in parallel."

2. "The team customized a BERT model to identify methodological weaknesses across published papers, though this was separate from our literature review process."

3. "Our platform combines AI text summarization with human expert oversight to conduct comprehensive literature reviews twice as fast as traditional methods."

4. "We processed 5,000 medical journals using deep learning, but ultimately our literature review was conducted through conventional critical appraisal techniques."

5. "The research assistant built an ML classifier that predicts which papers a researcher might want to include in their literature review based on previous selections."

6. "We're using AI to extract study characteristics from papers we've already identified, but the actual literature search and selection was done manually following PRISMA guidelines."

7. "Our software uses natural language processing to generate personalized reading lists from academic databases, though it's primarily designed for teaching purposes rather than formal literature reviews."

8. "We employed a GPT approach for our systematic review, but we also leveraged a plagiarism module to ensure originality."

#### Example Output

{{
"1": {{"REASONING": "GPT-4 used directly for systematic review synthesis",
       "SCORE": 9}},
"2": {{"REASONING": "AI for paper analysis, not literature review",
       "SCORE": 3}},
"3": {{"REASONING": "AI summarizes texts for literature reviews",
       "SCORE": 10}},
"4": {{"REASONING": "AI only for processing, not review tasks",
       "SCORE": 4}},

```
    "5": {{"REASONING": "ML assists paper selection for literature review",
            "SCORE": 8}},
    "6": {{"REASONING": "AI extracts data from already-selected papers",
            "SCORE": 6}},
    "7": {{"REASONING": "NLP for academic lists, not formal reviews",
            "SCORE": 5}},
    "8": {{"REASONING": "GPT for review but mentions plagiarism detection",
            "SCORE": 7}}
}}

## Task Instructions for **{tag}**

Now, evaluate the following statements for the code **{tag}** following the same procedure as above.
In some cases, I will also provide either (1) the question that the statement was given in response
to OR (2) text that immediately preceded the statement of interest (denoted with '...'). Remember you
are evaluating the STATEMENT itself - do not provide a high confidence score if the preceding
context/question is related to the code but the statement itself is not.

I may also provide additional context on the qualitative research dataset from which the following
statements extracted below. If provided, use this additional context to help guide your decision-
making.

{project_context}
**Code Name:**
{tag}
**Code Definition:**
{tag_definition} {tag_context}

**Inclusion Criteria:**
{inclusion_criteria}

**Exclusion Criteria:**
{exclusion_criteria}

**Tagged Examples (Relevant Statements)**:
{positive_statements}

**Untagged Examples (Irrelevant Statements)**:
{negative_statements}

#### Statements to Evaluate
{question_and_statements}
```

where *{tag}* is the name of the code being annotated, *{tag_defintion}* is the definition of this associated code, *{tag_context}* is an arrow diagram describing the tag's position within the codebook hierarchy, *{question_and_statements}* are the excerpts to be evaluated as well as their preceding their text provided as context, *{includsion_criteria}* and *{exclusion_criteria}* are the associated code's inclusion and exclusion criteria (respectively), *{project_context}* is a brief user-specified description of the dataset being analyzed, and *{positive_statements}* and *{negative_statements}* are user-specified few shot examples of both excerpts that should be coded with the associated code as well as those that should not, respectively.

*A.4.2 Multi-Code Application*

The prompt text leveraged for the *distribute_code* function is given below:

```
# Qualitative Research Task

You are an assistant to a qualitative researcher. Your task is to assign each statement from a
qualitative dataset to subthemes based on the provided parent theme and subtheme definitions. If a
statement does not clearly match any of the subthemes, respond with [].

---

## Instructions

1. **Input Data:**
- **PARENT THEME:** The overarching topic discussed within the statements
- **SUBTHEMES:** A list of subthemes, each identified by an alphabetic identifier (e.g., `A. Positive
sentiment`).
- **STATEMENTS:** A set of statements from the dataset, each identified by a numeric identifier. In
some cases, I will also provide either (1) the question that te statement was given in response to OR
(2) text that immediately preceded the statement of interest (denoted with '...'). Do not apply a
code if the preceding context/question is related to a code but the statement itself is not.

2. **Assignment Rules:**
- **Flexible Assignment:** Each statement can be assigned to **one, multiple, or no** subthemes.
- **Default Assignment:** If none of the provided subthemes clearly match a statement, assign [] to
that statement.
- **Strict Criteria:** Only assign a subtheme if the statement clearly meets the subtheme's definition,
inclusion criteria, and any provided examples.

3. **Output Format:**
- Your response should be in JSON format.
- Each key should be the numeric identifier of a statement (STATEMENT X formatted as `"X"`).
- Each value should be a list of the alphabetic identifier(s) of the assigned subtheme(s) (e.g.,
`["A"]`) or []if no subthemes apply.

---

## Example

**Input:**

- **PARENT THEME:**
Social Media Posts About the Election

- **SUBTHEMES:**
- A. Positive sentiment
- B. Negative sentiment

- **STATEMENTS:**
1. I am so anxious and worried about the election
2. I can't wait for the election to come because I'm excited to vote!
3. Thinking about the election is making me depressed
4. I'm really scared there is going to be another uprising again and people will get hurt
```

```
5. I'm feeling indifferent about the election. Given the candidates running, there really is much
difference between them in terms of how they are going to change the country.

**Expected Output:**

{{
"1": ["B"],
"2"  ["A"],
"3": ["B"],
"4": ["B","C"],
"5": []
}}

**Explanation:**

- 1: Clearly aligns with negative sentiment (SUBTHEME B)
- 2: Clearly aligns with positive sentiment (SUBTHEME A)
- 3: Clearly aligns with negative sentiment (SUBTHEME B)
- 4: Associated with both negative sentiment (SUBTHEME B) and statements about concern for violence
(SUBTHEME C)
- 5: Has a neutral tone and does not clearly match any subtheme, so it is assigned the default value
of [].

---

## Additional Information

At times, you may be provided with additional context on the qualitative research project from which
the statements are extracted as well as further details for each subtheme (e.g., definitions,
inclusion/exclusion criteria, positive and negative examples). Use this information to guide your
assignment decisions.

## Task
Now, perform this task for the inputs below:

{project_context}

PARENT THEME:
{theme}

SUBTHEMES:
{subthemes}

STATEMENTS:
{statements}

Please ensure your output is valid JSON with each statement key correctly mapped to a list of its
assigned subtheme alphabetic identifier(s).
```

where *{statements}* are the set of statements to be evaluated, *{subthemes}* are the set of subcodes associated with the distribute operation along with their definitions, *{theme}* is the parent code of the associated subcodes along with its definition (if available), and *{project_context}* is a brief user-specified description of the dataset being analyzed.

The above prompts are the foundational prompts for Muse's annotation features; however slightly modified prompts are used under different coding conditions (i.e. when *force_assign* is required by a distribute operation, when code definitions are not present for an apply operation, etc.).

## A.5 Error Analysis Discussion

Code under-specification often took the form of target codes with definitions that vaguely described the concept of interest. For example, the label "relationships" is defined as "family and social relationships," but lacked sufficient detail and criteria to accurately apply (or not apply) the code to Reddit posts about social interactions, relationship types, and community engagement. Adding specific information in the definition that the human coders implicitly or explicitly utilized when coding, such as "mention of family members, relatives, friends, significant others, or social roles," would have provided the LLM with much-needed context to improve alignment with the ground truth coding scheme.

The consequence of code under-specification was often codebook confusion. Codebook confusion does not imply either LLM or human correctness but rather is a feature of a given codebook specification. Codebook confusion includes cases when 1) a FP was assigned but another ground truth positive code is semantically similar to the FP such that the difference is ambiguous and/or 2) a codebook has implicit or explicit application criteria for certain codes like not allowing multi-codes between them or requiring at least one to be applied. The first case, ambiguity between semantically similar codes, was more prevalent than the second in this dataset because multi-coding and force-assign restrictions or requirements were not explicitly specified. We found that 25 of 49 FPs contained "codebook confusion." The most frequent example of such confusion in the sample was between the code "thinspiration," defined as "drive for thinness, want to be thinner or skinny," and the code "weight loss," defined as "body weight loss." An excerpt from an exemplar post containing this unresolved ambiguity is "I have a lot of thigh fat on me. I really want to start wearing shorts next year, but I want to look good wearing them, so I'm wanting to lose my thigh fat," which was coded positively for "thinspiration" but not for "weight loss" by the human coders. While some of these errors can be the result of insufficient codebook context and may be rectified if the full codebook is provided, it is also likely that some are incorrectly coded in the ground truth data due to coder fatigue and high code density.

A small number of errors by the LLM were inexplicably erroneous. One FP was a post about an individual suffering from an eating disorder that has led to weight loss but was coded by the LLM as restricting nutrition intake. Two FNs were the absence of the code "body dissatisfaction" when posts were clearly self-deprecating with respect to physical appearance, and one FN was about restricting carbohydrate intake on a high-caloric diet which was not coded positively for restricting nutrition intake. It's difficult to generalize based on these four cases.

Finally, 18 of 49 FNs and 6 of 49 FPs contained annotation inconsistencies. For example, a FN post envious of others' strength and well-developed muscles, saying "Today I hit 885lbs on the Big 3 […] I always thought I'd be looking 'jacked' to hit those types of numbers […] #ForeverSmall," was coded as "thinspiration" (drive for thinness) by the human coders while the authors interpreted it as actually about a drive for muscle development. Many of the other FNs that we determined to contain inconsistencies contained keyword matches with the target code in the document (e.g., "calories" for the code "calorie count" or "fitness" for the code "physical exercise") but were about a separate subject matter unrelated to the target code. A FP, on the other hand as an example, discusses an abusive relationship/conflict regarding food without specifying who the relationship is with (just using the pronoun "she"), but isn't coded positively by the human coders for "relationships."